\documentclass[twocolumn]{aastex63} 

\usepackage{graphicx}
\usepackage{dcolumn}
\usepackage{bm}
\usepackage{amsmath}
\usepackage{longtable}
\usepackage{multirow}
\usepackage{array}
\usepackage{amssymb}

\shorttitle{TeV Emission Come from the GRB Prompt Phase}
\shortauthors{Wang et al.}

\begin{document}

\title{Implication of GRB 221009A: Can TeV Emission Come from the GRB Prompt Phase?}

\author[0000-0003-4976-4098]{Kai Wang}
\affiliation{Department of Astronomy, School of Physics, Huazhong University of Science and Technology, Wuhan 430074, China; \url{kaiwang@hust.edu.cn}}

\author{Zhi-Peng Ma}
\affiliation{Department of Astronomy, School of Physics, Huazhong University of Science and Technology, Wuhan 430074, China; \url{kaiwang@hust.edu.cn}}

\author[0000-0003-1576-0961]{Ruo-Yu Liu}
\affiliation{School of Astronomy and Space Science, Nanjing University, Xianlin Road 163, Nanjing 210023, China; \url{ryliu@nju.edu.cn}}

\author[0000-0002-5400-3261]{Yuan-Chuan Zou}
\affiliation{Department of Astronomy, School of Physics, Huazhong University of Science and Technology, Wuhan 430074, China; \url{kaiwang@hust.edu.cn}}

\author{Zhuo Li}
 \affiliation{Department of Astronomy, School of Physics, Peking University, Beijing 100871, China;
}
 \affiliation{Kavli Institute for Astronomy and Astrophysics, Peking University, Beijing 100871, China;}
 
\author{Zi-Gao Dai}
 \affiliation{Department of Astronomy, University of Science and Technology of China, Hefei 230026, China}

 
\begin{abstract}
Recently, the B.O.A.T. (``brightest of all time'') gamma-ray burst, dubbed GRB~221009A, was detected by various instruments. Unprecedentedly, the GRB presented very-high-energy (VHE, energy above 0.1\,TeV) gamma-ray emission with energy extending above 10\,TeV, as reported by the Large High Altitude Air Shower Observatory (LHAASO). We here demonstrate that the VHE and especially $>10$\,TeV emission may originate from the internal hadronic dissipation of the GRB, without the need of invoking any exotic processes as suggested by some previous studies. The possible prompt origin of LHAASO photons may imply the first detection of the GRB prompt phase in the VHE regime. We also discuss the constraints on the properties of the GRB ejecta from multiwavelength and multi-messenger observations, which favors a magnetically dominated GRB ejecta. The suggested Poynting-flux-dominated GRB ejecta in this work supports the Blandford $\&$ Znajek (BZ) mechanism as the possible central engine model of GRB, as well as the possible strong magnetic dissipation and acceleration.

\end{abstract}


\section{Introduction}
High-energy gamma-rays ($>100\,\rm MeV$) have been found in gamma-ray bursts (GRBs), the most energetic explosions in the universe, by the Large Area Telescope (LAT) onboard the {\it Fermi} satellite~\citep{2013ApJS..209...11A,2019ApJ...878...52A,2021ApJ...922..255T}. However, the radiation mechanism of high-energy gamma-rays is still under debate. Currently, the origins of high-energy gamma-rays can be classified as two kinds of radiation mechanisms, i.e., the leptonic and hadronic origins. The former one is usually related to the synchrotron radiation or the inverse Compton (IC) scattering of the low-energy photon field through energetic electrons accelerated by the shocks ~\citep{2009MNRAS.400L..75K,2008MNRAS.385.1461Y,2010MNRAS.409..226K,2009A&A...498..677B,2011ApJ...739..103A,2011ApJ...729..114A,2012ApJ...757..115A,2013ApJ...773L..20L,2013ApJ...771L..33W,2014ApJ...788...36B,2017ApJ...844...92F}. In contrast, the hadronic origin of high-energy gamma-rays is caused by the accelerated protons, which can interact with the GRB's intense keV/MeV radiation field via hadronic processes, e.g., the photomeson production process ($p\gamma\to (p/n)\pi^0\pi^+\pi^-$) and Bethe-Heitler process (BH, $p\gamma\to pe^+e^-$). The secondary high-energy photons and electrons produced from hadronic processes will inevitably initiate the electromagnetic (EM) cascade via the $\gamma\gamma$ annihilation for high-energy photons and the synchrotron and IC process for high-energy electrons in the GRB environment, contributing to the observed high-energy gamma-rays~\citep{2009ApJ...705L.191A,2010ApJ...725L.121A,2012ApJ...746..164M,2012ApJ...757..115A,2018ApJ...857...24W,2022arXiv221200765R}.

Hadronic processes are generally suggested to occur if the charged nuclei can be accelerated to be ultra-high-energy cosmic rays (UHECRs). GRBs are thought to be the promising candidates to accelerate particles to ultrahigh energies \citep{1995PhRvL..75..386W,1995ApJ...453..883V,2010PhRvD..82d3008M}. However, the expected accompanying neutrinos produced by the hadronic processes have not been observed by IceCube, and consequently, the strong constraints on the GRB model parameters based on the combination of these parameters, such as the energy dissipation radius, the bulk Lorentz factor of the GRB jet, and the baryonic loading factor, have been achieved \citep{2012ApJ...752...29H,2013PhRvL.110l1101Z,2013ApJ...766...73L,2013ApJ...770L..40L,2015ApJ...805L...5A,2017ApJ...843..112A}.

In addition to high-energy gamma-rays, in recent years, Very-high-energy (VHE) gamma-rays have been detected in the afterglow phase by some VHE gamma-ray detectors, e.g., GRB 190114C~\citep{2019Natur.575..455M,2019Natur.575..459M} and GRB 201216C~\citep{2022icrc.confE.788F} by The Major Atmospheric Gamma Imaging Cherenkov (MAGIC) observatory, GRB 180720B~\citep{2019Natur.575..464A} and GRB 190829A~\citep{2021Sci...372.1081H} by the High Energy Stereoscopic System (HESS) observatory. These VHE gamma-rays with energies above $100\,\rm GeV$ but below $1\,\rm TeV$ have the same debate about their origins (see, e.g., \citep{2022Galax..10...74G}).

Recently, an extraordinarily bright and energetic GRB, GRB~221009A, triggered the Fermi Gamma-ray Burst Monitor (GBM) at $T_0=13:16:59.000 \,\rm UT$ on 9 October 2022~\citep{2022GCN.32636....1V,2022GCN.32642....1L}. and many other instruments, e.g.,  Fermi-LAT~\citep{2022GCN.32637....1B,2022GCN.32658....1P}, Swift~\citep{2022GCN.32688....1K,2022GCN.32632....1D}, Gravitational wave high-energy Electromagnetic Counterpart All-sky Monitor (GECAM)~\citep{2022GCN.32751....1L}, AGILE/MCAL~\citep{2022GCN.32650....1U}, Konus-Wind~\citep{2022GCN.32668....1F}. Some useful constraints on the GRB model have been obtained either by the neutrino non-detection from IceCube~\citep{2022arXiv221014116A,2022ApJ...941L..10M,2023arXiv230205459A} or by the Fermi-LAT measurement \citep{2023ApJ...943L...2L}. Especially, for the first time, the GRB was also captured by the extensive air shower detector, the Large High Altitude Air Shower Observatory (LHAASO)~\citep{2022GCN.32677....1H}, at the VHE band. Thanks to its high sensitivity, LHAASO recorded a huge amount of photons above $500\,\rm GeV$ from the GRB, and surprisingly discovered the emission above 10\,TeV from GRB for the first time. The origin of $>10\,\rm TeV$ photons has been attributed to the possible axion-like particles (ALPs) [e.g., \cite{2022arXiv221009250T,2022arXiv221007172B}]. Astrophysical processes of external origin for these VHE photons such as GRB afterglow's emission \citep{2022arXiv221010673R,2022arXiv221105754Z,2022arXiv221209266S,2023ApJ...942L..30S} or the EM cascade initiated by escaping ultrahigh-energy cosmic rays (UHECRs) in the intergalactic space have been also explored  \citep{2022arXiv221012855A,2022arXiv221013349D}. \cite{2022arXiv221200766R} studied the prompt emission within the internal shock scenario, considering  synchrotron radiation and the IC scattering of electrons, as well as the possible hadronic contribution. They also ascribed $>10\,\rm TeV$ photons to the EM cascade initiated by UHECRs in extragalactic background light (EBL). In this work, we aim to explore the internal origin of the VHE emission of GRB 221009A with paying a particular focus on whether $>10$\,TeV photon can possibly arise from the internal dissipation of the GRB. We will take into account the observational constraints from other instruments such as Fermi-LAT and IceCube, and explore available ranges of physical parameters.

The paper is organized as follows. The leptonic and hadronic models are described in Section~\ref{sec2}. Then we apply the models to GRB 221009A and the corresponding constraints are obtained in Section~\ref{sec3}. Finally, the conclusions and discussions are provided in Section~\ref{sec4}.

\section{Descriptions of Leptonic and Hadronic Models}\label{sec2}

We consider an isotropically expanding shell with the bulk Lorentz factor $\Gamma$ at radius $R=2 \Gamma^2 c \delta t$ from the central engine for a GRB with a variability timescale $\delta t$. The spectrum of keV/MeV photons in the prompt emission phase can be usually depicted by a broken-power-law distribution, i.e., $dn_\gamma/d\varepsilon_\gamma=A_\gamma(\varepsilon_\gamma/\varepsilon_{\gamma,p})^{q_\gamma}$ with a peak energy $\varepsilon_{\gamma,p}$, a low-energy index $q_\gamma=\alpha_\gamma$ for $\varepsilon_\gamma<\varepsilon_{\gamma,p}$ and a high-energy photon index $q_\gamma=\beta_\gamma$ for $\varepsilon_\gamma>\varepsilon_{\gamma,p}$. The normalized coefficient is $A_\gamma = {\Gamma^2U_\gamma }/\left[ {\int_{{\varepsilon _{\gamma,\min }}}^{{\varepsilon _{\gamma,\max }}} {(\varepsilon_\gamma/\varepsilon_{\gamma,p})^{q}{\varepsilon _\gamma }d{\varepsilon _\gamma }} } \right]$, where $U_\gamma =L_\gamma /(4 \pi R^2 \Gamma^2 c)$ is the photon energy density in the comoving frame, and $L_\gamma$ is the luminosity integrated from $\varepsilon _{\gamma,\min }$ to $\varepsilon _{\gamma,\max}$, which are fixed to be $1\,\rm keV$ and $10\,\rm MeV$ for calibration respectively. Although the radiation mechanism of the prompt keV/MeV radiations is not totally determined so far, e.g., the photospheric emission~\citep{2011ApJ...732...49P,2013ApJ...765..103L,2013MNRAS.428.2430L}, the Comptonized quasi-thermal emission from the photosphere~\citep{2005ApJ...628..847R,2014ApJ...785..112D}, the synchrotron emission of non-thermal electrons and so on (see a review, e.g., \cite{2014IJMPD..2330002Z}), we here employ the latter one, namely, the synchrotron emission of non-thermal electrons, to study. In order to explain the observed keV/MeV photons, accelerated non-thermal electrons with a broken-power-law distribution are introduced, i.e., $dn_e/d\gamma_e=A_e(\gamma_e/\gamma_{e,b})^{q_e}$with a break electron Lorentz factor $\gamma_{e,b}$, a low-energy index $q_e=\alpha_e$ for $\gamma_e<\gamma_{e,b}$ and a high-energy electron index $q_e=\beta_e$ for $\gamma_e>\gamma_{e,b}$. The acceleration (or energy dissipation) mechanisms could be by shocks or magnetic reconnections~\citep{1994MNRAS.270..480T,2011ApJ...726...90Z} accounting for the conversion from the energy of the GRB jet to the non-thermal energies of emitting particles. In our calculations, $\alpha_e$, $\beta_e$, $\gamma_{e,b}$, and $A_e$ are obtained based on the phenomenological spectral fittings to the observed keV/MeV photons. Especially, we calculate the electron energy density $U_e$ by integrating the electron distribution after the above parameters are determined and find the ratio of $U_e/U_\gamma \sim 1$ as the same as the fast-cooling regime that our cases are.

In addition, the primary protons are assumed to be accelerated to a power-law distribution in the GRB outflow, i.e., $d{n_p}/d\gamma _p = {A_p}{\gamma _p}^s$ for $\gamma_{p,\min} \leqslant \gamma_{p} \leqslant \gamma_{p,\max}$ \footnote{We neglected the possible exponential cutoff at the high-energy tail of proton distribution, i.e., $\exp{(-\gamma_{p}/\gamma_{p,\max})}$, because its impact is quite tiny for the proton distribution with a very broad energy range.}, where $\gamma_{p,\min}$ is taken to be just slightly larger than unity in the comoving frame and $\gamma_{p,\max}$ is determined by the balance between the acceleration timescale and the cooling timescale (or the dynamical timescale), namely, ${t_{\rm acc}} = \min \{ {t_{\rm cooling}},{t_{\rm dyn}}\} $.  The dynamical timescale in the comoving frame is $t_{\rm dyn}\simeq R/\Gamma c$. The comoving acceleration timescale is ${{t}_{\rm acc}} \simeq \eta {{\gamma }_p}{m_p}c/eB$ in the magnetic field strength $B$ with the electron charge $e$ and the Bohm factor $\eta(\ge 1)$ which indicates the deviation from the acceleration in the Bohm limit. In this work we adopt the Bohm diffusion ($\eta=1$) under the assumption that the Larmor radius equals the correlation length of the magnetic field. The realistic acceleration may deviate from the Bohm diffusion~\citep{2005ApJ...627..868G,2014A&A...569A..58W}, inducing a larger Bohm factor. A larger Bohm factor $\eta$ will result in a smaller maximum proton energy, and further, affect the normalization factor $A_p$. However, the impact on $A_p$, as well as the subsequent emission, is basically small for a general flat ($s\simeq-2$) proton distribution (e.g., by the Fermi acceleration). Besides, around the maximum proton energy, the hadronic processes usually have relatively high interaction efficiency, so a smaller maximum proton energy will reduce the flux level of the cascade emission to some extent. The considered cooling processes for protons are synchrotron radiation, the photomeson production process, and the BH process. The comoving synchrotron cooling timescale for the relativistic proton is $t_{\rm syn} = \frac{9(\gamma_p-1)}{4{\gamma_p} ^2}\frac{{m_p}^3 c^5}{e^4B^2}$. The photomeson production and BH timescales are calculated by integrating their productions following the semi-analytical treatment suggested in~\cite{2008PhRvD..78c4013K}. The baryonic loading factor is obtained by the ratio between the energy density of accelerated protons in the comoving frame and that of keV/MeV photons, say, $f_p\equiv U_p/U_\gamma$. The magnetic energy density is achieved by introducing a factor $f_B\equiv U_B/U_\gamma$, and consequently, the magnetic field strength in the comoving frame can be written as $B=\sqrt{8\pi U_B}=\sqrt{2f_B L_\gamma/\Gamma^2R^2c}$.

The keV/MeV photons can be described by the synchrotron radiation of primary electrons. The high-energy gamma-rays with energies above $100\,\rm MeV$ that in some GRBs (e.g., GRB 090902B~\citep{2009ApJ...706L.138A}, GRB 090926A~\citep{2011ApJ...729..114A}, and the GRBs listed \cite{2021ApJ...922..255T}) can be shown as a distinct spectral shape (Note that GRB 221009A is this case based on the observations described in Section~\ref{sec:observations}) can be ascribed to leptonic or hadronic processes. We refer to both as the \textit{lepton-dominated} scenario and the \textit{hadron-dominated} scenario, respectively. For the lepton-dominated case, high-energy gamma-rays can be produced by the Self-synchrotron Compton (SSC) process of primary electrons and the subsequent EM cascade inside the GRB jet. While for the hadron-dominated case, the EM cascade initiated by the secondary photons and $e^\pm$ pairs of hadronic processes (including both photomeson production and BH processes) is responsible for the observed high-energy gamma-rays. In addition, for the proton-induced cascade, the secondary productions, e.g., electrons and neutrinos, will be suppressed since the intermediate particles such as charged pions and muons may cool down through the synchrotron radiation before they decay (see, e.g., \cite{2007PhRvD..75l3005L,2012JCAP...10..020B,2013ApJ...773..159B,2015JCAP...09..036T,2020PhRvD.102l3008B}). As a result, the suppression factors $1-\exp( - {{t}_{\pi,\rm syn}(E_\pi)}/{{\tau }_\pi }(E_\pi))$ and $1-\exp( - {{t}_{\mu,\rm syn}(E_\mu)}/{{\tau }_\mu(E_\mu) })$ due to the synchrotron cooling for charged pions and muons are respectively involved, where $E_\pi =0.2 E_p$ and $E_\mu =0.15 E_p$ are the energies of pions and muons relying on the parent proton energy, and ${\tau }_\pi =2.6 \times 10^{-8}\gamma_\pi\,\rm s$ and ${\tau }_\mu =2.2 \times 10^{-6}\gamma_\mu\,\rm s$ are the lifetimes of pions and muons. Our calculations are based on the conventional one-zone model, i.e., all physical processes occur in the same region, so the suppression factors are calculated by assuming the pions and muons cool down in the same dissipation region with the same magnetic field strength. The suppression factors are basically small for the typical magnetic field strength and may play a role only for the most energetic pions and muons in a strong magnetic field. As a result, the EeV neutrinos may be suppressed to some extent if a large magnetic field is involved but the influence on the PeV neutrino production is generally negligible. For simplicity, the synchrotron radiation of these intermediate charged pions and muons is neglected as their contribution to EM cascades is always sub-dominated considering the comparable generated neutral pions and charged pions.

\begin{figure}
	\centering
	\includegraphics[width =0.99\linewidth , trim = 25 5 55 30,clip]{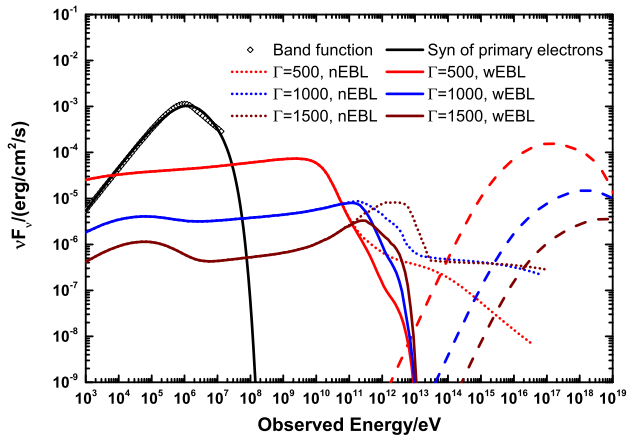}
	\caption{The spectra of synchrotron radiations of primary electrons (black solid line), total cascade emission with the EBL absorption (colored solid lines) and without the EBL absorption (colored dotted lines), and the produced all-flavor neutrino flux (corresponding dashed lines) for different bulk Lorentz factors. The observed keV/MeV radiation is substituted by the Band function (hollow diamond) with $\alpha_\gamma=-1.1$, $\beta_\gamma=-2.6$ and $\varepsilon_{\gamma,p}=1\,\rm MeV$. The total cascades contain the SSC-initiated and proton-initiated components. The adopted parameters are listed in Table~\ref{tab:table1}. To save the computation time, the outputs of cascade emissions during our numerical calculations cease at $100\,\rm TeV$ in the comoving frame, corresponding to $100\,(\Gamma/1000)/(1+z)\,\rm PeV$ in the observer's frame.}
	\label{fig1}
\end{figure}

Our treatment of the EM cascade process is implemented as detailed in the previous study \citep{2018ApJ...857...24W}. The observed spectral properties of keV/MeV radiations, i.e., the spectral indexes $\alpha_\gamma$ and $\beta_\gamma$, the peak energy $\varepsilon_{\gamma,p}$, and the luminosity $L_\gamma$ in 1\,keV$-$10\,MeV, are mainly ascribed to the distribution of primary electrons, namely, the electron distribution indexes $\alpha_e$ and $\beta_e$, and the break electron Lorentz factor $\gamma_{e,b}$. Other free parameters are the bulk Lorentz factor $\Gamma$, the variability timescale $\delta t$ (or the dissipation radius $R=2\Gamma^2 c \delta t$), the baryonic loading factor $f_p$, the magnetic energy fraction $f_B$, and the proton spectral index $s$. With these parameters, the keV/MeV photon field, the magnetic field, the electron distribution, and the proton distribution are determined. Consequently, the synchrotron radiation, the SSC radiation, and the photomeson production and BH processes can be calculated. Then SSC photons, secondary gamma-rays and electrons from hadronic processes are treated as the first-generation injection particles to participate in the EM cascade process (for the detailed treatment, see \cite{2018ApJ...857...24W}). In addition to target photons from synchrotron radiations of primary electrons, the EM cascade emission can also contribute as target photons to the photo-hadronic interactions. During the calculation of the photomeson production and BH processes, we directly used the observed keV/MeV radiations as the target photons since the final sum of synchrotron radiation of primary electrons and the cascade emission has to match the observations.

\begin{table}
\caption{\label{tab:table1}The Adopted Parameters in Fig.~\ref{fig1}.
}
\begin{ruledtabular}
\begin{tabular}{ccc}
	\textrm{Descriptions}&\textrm{Symbols}&\textrm{Values}\\ 
        \colrule
	        Redshift 	&$z$ 	& 0.15 \\ 
			Variability timescale 	&$\delta t$  & $0.082\,\rm s$  \\
			Dissipation radius 	&$R$  & $2\Gamma^2 c \delta t$  \\
			Low energy photon index 	&$\alpha_\gamma$  	& -1.1  \\
			High-energy photon index 	&$\beta_\gamma$  	& -2.6 \\
			Peak energy 	&$\varepsilon_{\gamma,p}$  	& $1\,\rm MeV$  \\
			Low energy electron index 	&$\alpha_e$  	& -1  \\
			High-energy electron index 	&$\beta_e$  	& -4.2 \\
			Proton index 	&$s$  	& -2  \\
			Calibration luminosity 	&$L_\gamma$\footnote{The luminosity at 1\,keV--10\,MeV.}&$2\times 10^{53}\,\rm erg/s$  \\
			Bulk Lorentz factor 	&$\Gamma$  	& [500,1000,1500]  \\
			Electron break Lorentz factor 	&$\gamma_{e,b}$ & [5210,10420,15630] \\
			
			Baryonic loading factor 	&$f_p$  & 10 \\
			Magnetic energy factor 	&$f_B$  & 1 \\
\end{tabular}
\end{ruledtabular}
\end{table}

\begin{figure}
	\centering
	\includegraphics[width =0.99\linewidth , trim = 25 5 55 30,clip]{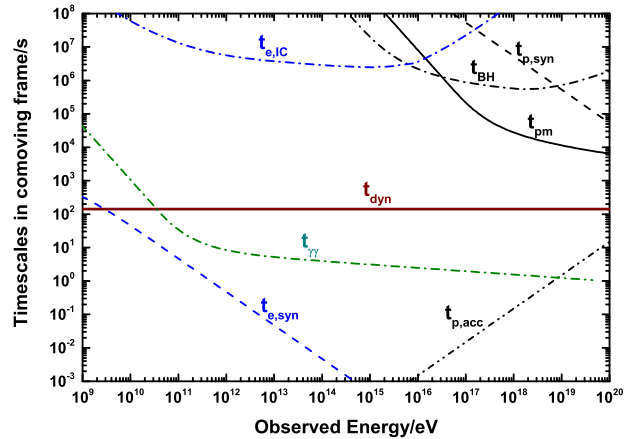}
	\caption{The timescales in the comoving frame for various processes. The bulk Lorentz factor $\Gamma=1000$ is adopted and other adopted parameters are the same as listed in Table~\ref{tab:table1}.}
	\label{fig2}
\end{figure}

\begin{figure}
	\centering
	\includegraphics[width =0.99\linewidth , trim = 25 5 55 30,clip]{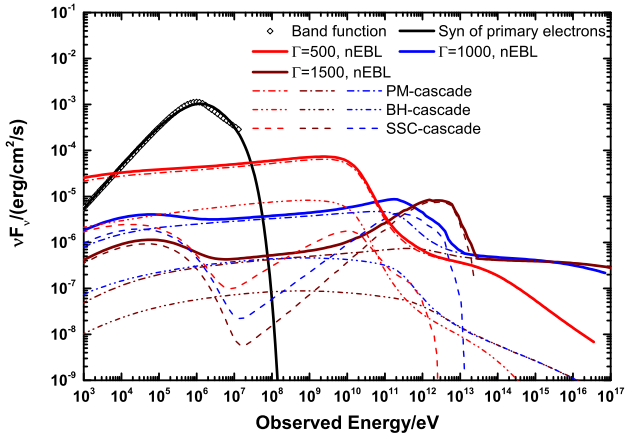}
	\caption{The spectra of synchrotron radiations of primary electrons (black solid line), total cascade emission without the EBL absorption (colored solid lines) for different bulk Lorentz factors, i.e., 500 (red), 1000 (blue), 1500 (wine). The total cascade emission includes the cascade emissions initiated by the secondaries of the photomeson production process (PM-cascade, dash-dotted lines), the secondaries of the BH process (BH-cascade, dash-dot-dotted lines), and the SSC photons (SSC-cascade, dashed lines). The adopted parameters are the same as in Fig.~\ref{fig1} and listed in Table~\ref{tab:table1}.}
	\label{fig5}
\end{figure}

The cascade spectra are presented in Fig.~\ref{fig1} under the parameter values listed in Table~\ref{tab:table1}, and the comoving timescales for various processes are shown in Fig.~\ref{fig2}. The baryonic loading factor $f_p=10$ and the magnetic energy factor $f_B=1$ are adopted as the benchmark values. Besides, the values of other parameters are adopted as the observations of GRB 221009A as introduced in Section~\ref{sec:observations}. The new EBL model given by \cite{2021MNRAS.507.5144S} is adopted for numerical calculations. Generally, a large bulk Lorentz factor induces a larger dissipation radius and a consequent smaller flux of cascade emission due to the smaller interaction efficiency for the smaller number densities of low-energy photons, electrons, and protons. Each component of total cascade emission is presented in Fig.~\ref{fig5}. Basically, the cascade spectrum initiated by the SSC photons is hard since it is dominated by the unabsorbed SSC photons that can keep a similar spectral shape to the synchrotron radiation, whereas the cascade emission initiated by the hadronic processes is generally flat and more or less universal as the cascade emission is fully developed. For a smaller dissipation radius, the internal $\gamma\gamma$ absorption inside the GRB jet becomes more dominant, inducing a smaller cutoff energy around GeV--TeV. In addition, the cascade emission becomes dominated by hadronic processes. For a large dissipation radius and other adopted parameter values, e.g., $\Gamma=1500$, the TeV photons are mainly contributed by the cascade process initiated by the SSC photons, whereas the GeV photons can be from the proton-initiated cascade emission, resulting in possible different radiation mechanisms between GeV photons and TeV photons.

\section{Application to GRB 221009A}\label{sec3}
\subsection{Observations of GRB 221009A}\label{sec:observations}
At $T_0=13:16:59.000 \,\rm UT$ ($T_0$) on 9 October 2022, GRB 221009A was triggered by the Fermi-GBM~\citep{2022GCN.32636....1V}. The estimated redshift for this GRB is $z= 0.151$ \citep{2022GCN.32648....1D}. GRB 221009A is a long-lasting GRB with a lower limit of $T_{90}$ (15-350 keV) is $\sim 1068\,\rm s$ \citep{2022GCN.32688....1K} and an extraordinarily bright and energetic GRB with a record-breaking fluence of $\sim 0.052 \,\rm erg/cm^{2}$ in the interval from $T_0$ to $T_0+600\,\rm s$ \citep{2022GCN.32668....1F}. The time-averaged spectrum at the onset of the brightest phase of this GRB prompt regime (from $T_0+180\,\rm s$ to $T_0+200\,\rm s$) is best fitted in the $20\,\mathrm{keV}-15\,\mathrm{MeV}$ energy range with the low-energy photon index $\alpha_\gamma=-1.09\pm 0.01$, the high energy photon index $\beta_\gamma=-2.60\pm 0.06$ and the peak energy $\varepsilon_{\gamma,p}=1060^{+31}_{-30}\,\rm keV$~\citep{2022GCN.32668....1F}. High-energy gamma-rays are detected by Fermi-LAT even extending for about 25 ks post GBM trigger and the highest-energy photon is $99.3\,\rm GeV$ observed 240 seconds after Fermi-GBM trigger \citep{2022GCN.32658....1P}. Besides, the estimated photon index above 100 MeV is $-1.87\pm0.04$ in the time interval $200-800\,\rm s$ \citep{2022GCN.32658....1P}.

Owing to the extreme brightness of GRB 221009A, most detectors are under the instrumental pile-up effects due to the data saturation in the main burst period (from $T_0+220\rm s$ to $T_0+270\rm s$) except GECAM \citep{2022GCN.32751....1L}. The gamma-ray detector GRD01 onboard GECAM records the maximum flux of $\sim 3\times 10^4$ counts/s for the main burst period of GRB 221009A lasting around several seconds from 400 keV to 6 MeV \citep{2022GCN.32751....1L}. The translated maximum flux for the main burst period can be conservatively estimated as $\sim 0.02\,\rm erg/cm^2/s$
lasting for around several seconds by considering the effective area of GRD01 at $400\,\rm keV$ is $\sim 1\,\rm cm^2$ \citep{2020SSPMA..50l9509G} and all received photons are with energies of $400\,\rm keV$.

LHAASO also reported the detection of $\sim 5000$ Very-High-Energy (VHE) photons ($>500\,\rm GeV$) within 2000 seconds after $T_0$, and the highest-energy photon is up to around 18 TeV \citep{2022GCN.32677....1H}. Moreover, IceCube neutrino observatory has carried out a search for track-like muon neutrino events arriving from the direction of GRB 221009A and derived a time-integrated muon-neutrino flux upper limit of $3.9\times10^{-2}\,\rm GeV/cm^2$ in the time interval from $T_0-1\,\rm hour$ to $T_0+2\,\rm hours$ under the assumption that the power law index of the neutrino distribution is $-2$ \citep{2022GCN.32665....1I}.

We consider two time intervals: the first one is $300-400\,\rm s$ since during this period the spectrum of Fermi-GBM and Fermi-LAT can be derived without the data saturation \citep{2023ApJ...943L...2L}, and the second one is $200-300\,\rm s$ which is the most energetic burst period. 

For the time interval $300-400\,\rm s$, the variability timescale is $\delta t=0.082\,\rm s$ \citep{2023ApJ...943L...2L}, yielding a dissipation radius $R=2 \Gamma^2 c \delta t \simeq 5 \times 10^{15}\,(\Gamma/1000)^2\,\rm cm$ \footnote{A large radius of $10^{16}-10^{17}\,\rm cm$ adopted in \cite{2022arXiv221200766R} is based on the preliminary data of INTEGRAL/SPI-ACS during the brightest emission period of the GRB, giving a long variability timescale 1.4\,s. However, as indicated by \cite{2022GCN.32660....1G}, the instrument is saturated during the peak of the GRB. Therefore, the short-scale structures in the lightcurve are likely smoothed out because of the saturation. We adopt the short-term temporal variability based on the standard Bayesian block method for the Fermi-GBM data before the brightest period of the event (i.e., before the saturation of GBM) obtained in \cite{2023ApJ...943L...2L}.}. The observed spectral properties of keV/MeV radiations, i.e., the low-energy photon index $\alpha_\gamma=-1.1$, the high energy photon index $\beta_\gamma=-2.6$ and the peak energy $\varepsilon_{\gamma,p}=1\,\rm MeV$ as the suggested spectral shape in \citep{2022GCN.32668....1F} are adopted. The peak flux of keV/MeV radiations is adopted as $\sim 10^{-4}\,\rm erg/cm^2/s$ \citep{2023ApJ...943L...2L}. For the time interval $200-300\,\rm s$, we adopt the same variability timescale $\delta t=0.082\,\rm s$ as in the time interval $300-400\,\rm s$, which is reasonable as seen from the lightcurve produced by GECAM GRD01 \citep{2022GCN.32751....1L}. The observed spectral properties of keV/MeV radiations are also taken as in \cite{2022GCN.32668....1F}, i.e., $\alpha_\gamma=-1.1$, $\beta_\gamma=-2.6$ and $\varepsilon_{\gamma,p}=1\,\rm MeV$. As the above analyses, the averaged peak flux of $\sim 10^{-3}\,\rm erg/cm^2/s$ in 100 seconds interval is adopted as suggested by GECAM GRD01 as the pileup effect is negligible for this instrument.

\subsection{Results}\label{secresults}
Considering the possible radiation contribution by the external shock, in the prompt phase, one has some constraints as below: (1) The detection number of VHE photons ($>500\,\rm GeV$) in the prompt phase should be lower than the LHAASO detection number $\sim 5000$ within 2000 seconds; (2) The gamma-ray emission at the Fermi-LAT energy band should be lower than the observations; (3) The detection number of high-energy neutrinos should be lower than 3 since the probability of non-detection will be less than $5\%$ for $N_\nu >3$ given that the detection probability follows the Poisson distribution. 

We evaluate the expected (anti)muon neutrino event number based on the generated neutrino flux and the effective area of IceCube ($100\,\mathrm{GeV}-10\,\mathrm{EeV}$) for a point source at the declination of this GRB ($\delta=19.8^{\circ}$) \citep{2021arXiv210109836I} and the expected VHE photons ($>500\,\rm GeV$) based on the cascade emission and the effective area of LHAASO. The effective area of the LHAASO Water Cerenkov Detector Array (WCDA) for the zenith angle $\theta=15^{\circ }-30^{\circ}$ and that of the LHAASO larger air shower kilometer square area (KM2A) are derived from~\citep{2019arXiv190502773C}. The expected VHE photon number by LHAASO is calculated by
\begin{equation}
N( > {E_\gamma }) = \int_{{E_\gamma }}^{{E_{\gamma ,\max }}} {F({E_\gamma })} A_{eff}^\gamma ({E_\gamma },\theta )Td{E_\gamma },
\end{equation}
where $F({E_\gamma })$ is the GRB flux after the EBL absorption, $A_{eff}^\gamma ({E_\gamma },\theta )$ is the photon effective area including LHAASO-WCDA and LHAASO-KM2A, and $T=100\,\rm s$ is integration time for each time interval. Assuming the most energetic photon $\sim 18\,\rm TeV$ is detected by LHAASO-KM2A, the relative energy resolution of which at this energy band is $\simeq40\%$ \citep{2019arXiv190502773C}, in the following, we conservatively explore the detection number of LHAASO for photons with energies above $10\,\rm TeV$ instead of $18\,\rm TeV$. For the Fermi-LAT data, the analyzed spectrum for $294-400\,\rm s$ by \cite{2023ApJ...943L...2L} is involved for the time interval $300-400\,\rm s$, which is an approximately power-law spectral shape with a photon index $-1.87\pm0.04$ and a peak flux of $\sim 10^{-5}\,\rm erg/cm^2/s$ \citep{2022arXiv221105754Z}. For the time interval $200-300\,\rm s$, we consider a similar spectral shape with a photon index $-1.87\pm0.04$ detected by Fermi-LAT as in the time interval $300-400\,\rm s$ but a larger peak flux of $\sim 10^{-4}\,\rm erg/cm^2/s$ as shown in \cite{2023ApJ...943L...2L}.

\subsubsection{Hadronic Constraints}\label{sec:hadconstraints}
We numerically calculate the spectra of synchrotron radiations of primary electrons, the cascade emission initiated by the SSC photons, secondary photons and electrons of the photomeson production process, and the secondary electrons of the BH process. The keV/MeV observations are explained by the synchrotron of primary electrons and the required electron distribution index can be easily obtained, $\alpha_e \simeq 2\alpha_\gamma+1 =-1.2$ ($\alpha_e=-1$ is used in the actual numerical calculation) and $\beta_e \simeq 2\beta_\gamma+1 =-4.2$. The electron distribution indexes are derived by phenomenological spectral fittings. The steady-state high-energy electron index $\beta_e$ usually can be easily obtained by an accelerated electron injection with an index of $\beta_e+1$ for the standard synchrotron radiation cooling, whereas the low-energy electron index in the standard synchrotron fast cooling regime is $-2$, corresponding to an observed low-energy photon index $\simeq -1.5$ softer than that in GRB 221009A. For the fast-cooling synchrotron radiation in the internal shock scenario, such a hard observed low-energy photon index generally needs to invoke the possible evolutional magnetic field in the post-shock region~\citep{2014NatPh..10..351U,2021Galax...9...68W}. Besides, the particle acceleration by the magnetic reconnection scenario can also solve the low-energy spectral index issue to some extent~\citep{2011ApJ...726...90Z}. Moreover, based on the Fermi-LAT observations (a single power law with a photon index of $\simeq -1.87$~\citep{2022GCN.32658....1P,2023ApJ...943L...2L}), the gamma-rays with energies above 100 MeV show a distinct spectral component from the GBM observations (a broken power law with a high energy photon index $\simeq -2.6$ above 1 MeV~\citep{2022GCN.32668....1F}). We treat the GBM observations and Fermi-LAT observations as two spectral components with different origins. GBM observations have been ascribed to the synchrotron radiation of primary electrons with a broken power law distribution. In addition, the flux level of the Fermi-LAT observation is much lower than that of the GBM observation, a too large synchrotron high-energy cutoff energy would violate the Fermi-LAT observations. Therefore, the maximum emission energies of synchrotron radiations of primary electrons are limited to be lower than $\sim 100\,\rm MeV$.

\begin{figure}
	\centering
	\includegraphics[width =0.99\linewidth , trim = 25 5 55 30,clip]{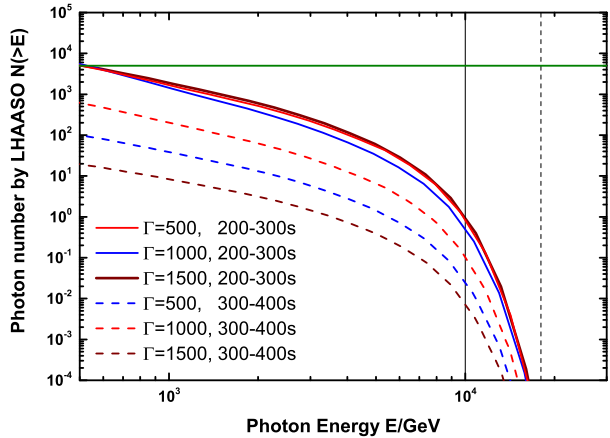}
	\caption{The detection number of VHE photons of hadronic origin by LHAASO. The contribution to the LHAASO detection number is limited as $N_\gamma(>500\,\mathrm{GeV}) \le 5000$ (olive horizontal line). The black vertical solid and dashed lines indicate the photon energy of $10\,\rm TeV$ and $18\,\rm TeV$, respectively. For both time intervals, the same $f_p$ is adopted, say, [2, 0.83, 1.86] for $\Gamma=$ [500, 1000, 1500]. The adopted luminosity at 1\,keV--10\,MeV is $2\times 10^{53}\,\rm erg/s$ for $200-300\,\rm s$ and $2\times 10^{52}\,\rm erg/s$ for $300-400\,\rm s$. The other parameters are the same as in Table~\ref{tab:table1}. The results are concluded in Table~\ref{tab:table2}.}
	\label{fig3}
\end{figure}

\begin{figure}
	\centering
	\includegraphics[width =0.99\linewidth , trim = 25 5 55 30,clip]{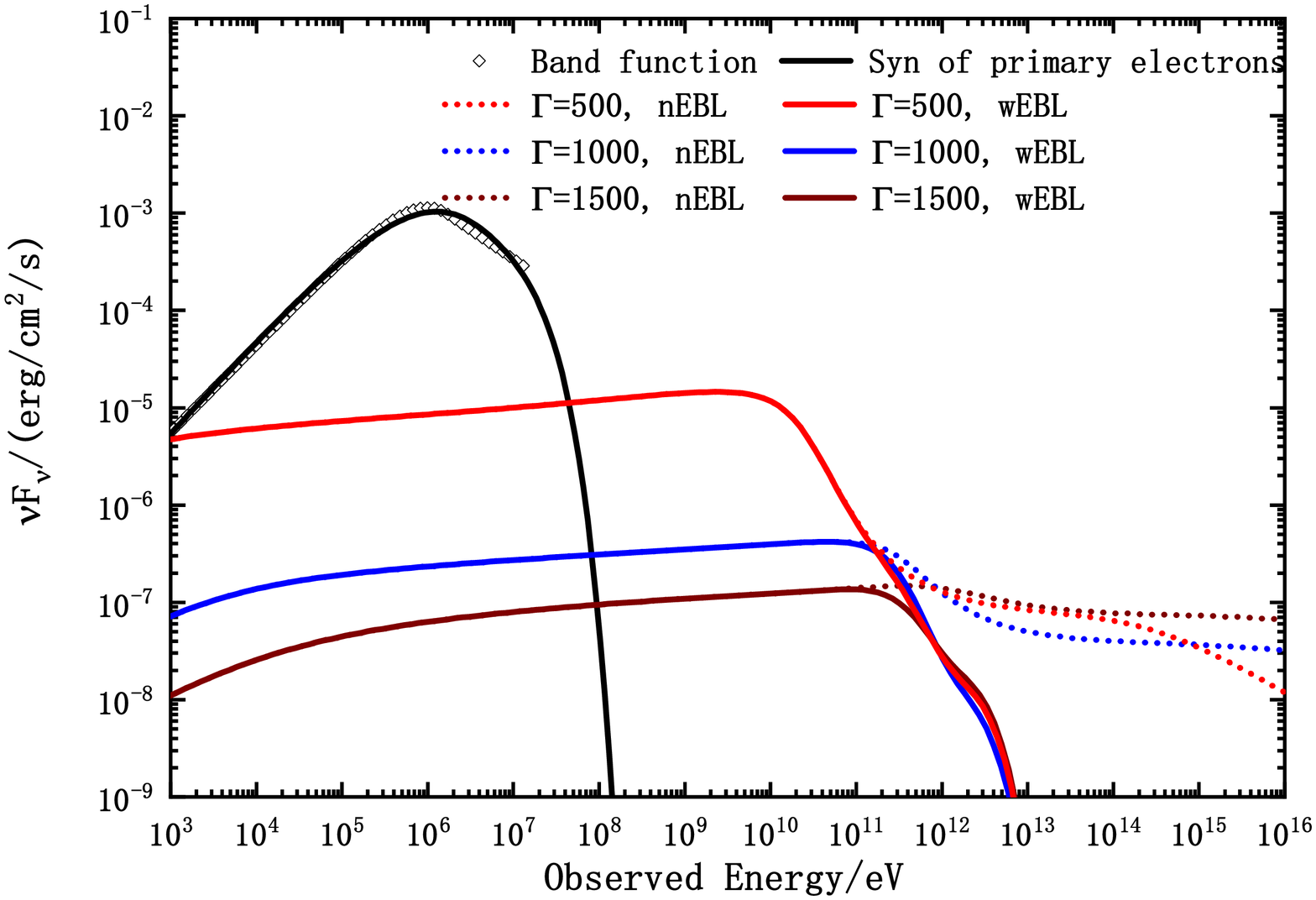}
	\caption{The corresponding spectra for hadronic constraints in the time interval 200--300\,s. The contribution of the SSC component is neglected. The adopted parameters are the same as in Fig.~\ref{fig3}.}
	\label{fig6}
\end{figure}

\begin{table}
\caption{\label{tab:table2}Constraints on the hadronic component under the assumption that the detection number by LHAASO is $\le 5000$ above 500\,GeV for the total contribution of two prompt time intervals.
}
\doublerulesep 0.1pt \tabcolsep 0.01pt
\begin{ruledtabular}
\begin{tabular}{ccc}
	\textrm{Descriptions}&\textrm{Symbols}&\textrm{Values}\\ 
        \colrule
			Bulk Lorentz factor 	&$\Gamma$  	& [500, 1000, 1500]  \\	
			Baryonic loading factor 	&$f_p$  & $\le$[2, 0.83, 1.86] \\
			Neutrino number\footnote{in 100\,GeV--10\,EeV.} 	&$N_\nu$  & $\le$[0.74, $5\times10^{-3}$, $9\times10^{-4}$] \\
            VHE photon number ($>10\,\rm TeV$) & $N_\gamma$  & $\le$[0.8, 0.6, 0.8]
\end{tabular}
\end{ruledtabular}
\end{table}

In the hadronic scenario, the SSC component of electrons is neglected, which will induce conservative hadronic constraints considering the contribution of the SSC component. The constraint given by the LHAASO detection is generally much more dominant than that given by the Fermi-LAT and high-energy neutrino observations. We normalize the VHE photon ($> 500\,\rm GeV$) detection number to $5000$ in Fig.~\ref{fig3}. Under the constraint of VHE photon detection ($> 500\,\rm GeV$) number by LHAASO $\le 5000$, we obtain the upper limit of the baryonic loading factor $f_p$. For the different bulk Lorentz factors, the required baryonic loading factor is $f_p \lesssim 2$, which is much stronger than that obtained by the constraints of high-energy neutrinos, especially for large bulk Lorentz factors (see, e.g., \cite{2022ApJ...941L..10M}). The corresponding spectra for the dominant time interval, i.e., 200--300\,s, are also presented in Fig.~\ref{fig6} with the same parameters as in Fig.~\ref{fig3}. As we can see, for a larger $\Gamma$, the cascade emission can be lower around GeV, whereas, around the TeV energy band, the intrinsic cascade emission (without EBL absorption) can be comparable with the case with the smaller $\Gamma$ due to the smaller internal $\gamma\gamma$ absorption inside the GRB jet. This generates almost the same limitations on the baryonic loading factor. The expected numbers of high-energy muon and antimuon neutrino event and $\gtrsim10\,\rm TeV$ VHE photon are listed in Table~\ref{tab:table2}. The expected $\nu_\mu + \nu_{\bar{\mu}}$ neutrino event number is basically small and the detection number of $\gtrsim10\,\rm TeV$ photon can be around unity. It suggests the sub-TeV and multi-TeV photons can be produced in the GRB prompt phase and the constraints given by the LHAASO observations can be more efficient than that given by the neutrino observations for the nearby GRB source.

Note that the proton spectral index $s=-2$ suggested by the general Fermi acceleration is adopted in our calculations. Deviation of the proton spectrum from $-2$ will not affect the spectral shape of the cascade emission as long as the EM cascade is fully developed. The fully-developed EM cascade can be seen in Fig.~\ref{fig6}, which shows a universal flat cascade spectrum. However, a softer proton spectrum would relax the constraint on $f_p$ as more energies would be carried by low-energy protons which have low efficiency of pion production. On the other hand, the spectral shape of produced high-energy neutrinos can be affected by the proton spectral index since they are produced directly from protons. However, since the detected neutrino number listed in Table~\ref{tab:table2} is basically very small, the change in the predicted neutrino number will not be significant enough to violate the non-detection of neutrinos.

In addition, a typical $f_B=1$ is used for hadronic constraints. Since we implement the multi-wavelength spectral constraints, during the hadronic constraints, $f_B$ is introduced to explain the keV/MeV radiations with the primary electrons and also be used for the EM cascade calculation. Different $f_B$ will not affect the low-energy keV/MeV photon number density which is determined by the observed keV/MeV radiations, bulk Lorentz factor, and the dissipation radius. The cascade emission initiated by secondary particles of hadronic processes can be fully developed by the synchrotron radiation, the inverse Compton, and the electron pair production. For the diverse $f_B$, the hadron-initiated cascade emission can be affected slightly (see Figure 6 in~\cite{2018ApJ...857...24W}). For a very low magnetic field, the cascade emission can be dominated by the inverse Compton of steady-state cascaded electrons, showing a slightly different spectral index but a comparable flux at the high-energy band. For a relatively large magnetic field (e.g., $f_B>0.1$), the cascade emission will be dominated by the synchrotron radiation of steady-state fully-developed cascaded electrons, showing a flat spectral shape and almost the same flux.

The SSC component with a small $f_B$ may violate the observations and in principle, a larger $f_B$ has been involved to lower its contribution (see Section~\ref{Sec:LepConstraints}). However, the calculations of the precise contribution of the SSC component with the diverse $f_B$ to the LHAASO and Fermi-LAT observations make our constraints complicated and unintuitive when we carry out the hadronic constraints. Besides, the impact of $f_B$ on the hadron-initiated cascade emission is weak, so we can implement relatively independent and conservative constraints on $f_p$ by assuming the hadronic contribution to the LHAASO and Fermi-LAT energy band should be lower than observations whatever how large other contributions by other processes. Therefore, a typical $f_B=1$ is used in the hadronic constraints.

\subsubsection{Leptonic Constraints}\label{Sec:LepConstraints}

\begin{figure}
	\centering
	\includegraphics[width =0.99\linewidth , trim = 25 5 55 30,clip]{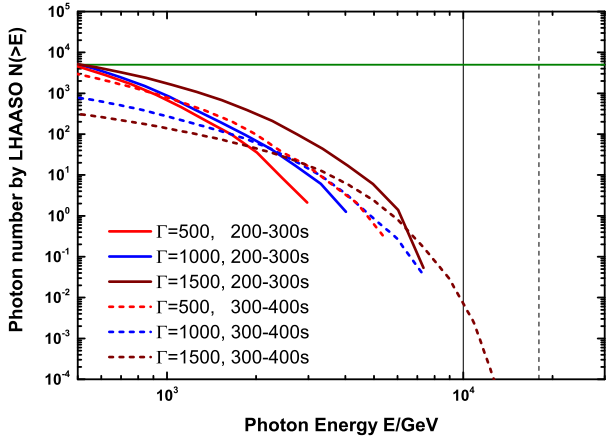}
	\caption{The detection number of VHE photons of leptonic origin by LHAASO. The meanings of lines are the same as in Fig.~\ref{fig3}. For both time intervals the same $f_B$ is adopted, say, [0.8, 50, 150] for $\Gamma=$ [500, 1000, 1500]. Besides, $f_p=0$ is adopted for both time intervals. The adopted luminosity at 1\,keV--10\,MeV is $2\times 10^{53}\,\rm erg/s$ for $200-300\,\rm s$ and $2\times 10^{52}\,\rm erg/s$ for $300-400\,\rm s$. The other parameters are the same as in Table~\ref{tab:table1}. The results are concluded in Table~\ref{tab:table3}.}
	\label{fig4}
\end{figure}

\begin{figure}
	\centering
	\includegraphics[width =0.99\linewidth , trim = 25 5 55 30,clip]{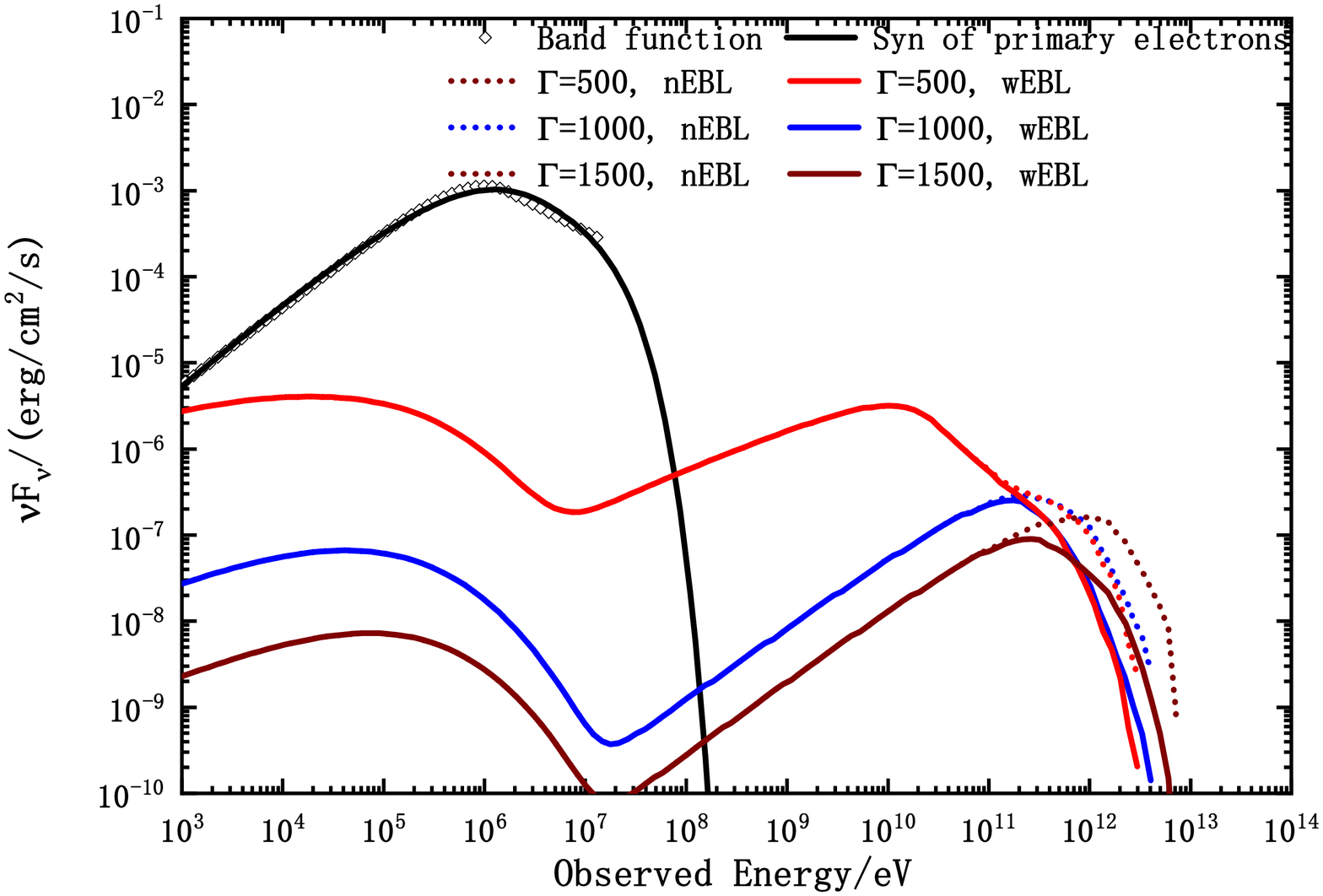}
	\caption{The corresponding spectra for leptonic constraints in the time interval 200--300\,s. The contribution of the hadronic component is neglected. The adopted parameters are the same as in Fig.~\ref{fig4}.}
	\label{fig7}
\end{figure}

We implement similar constraints on the lepton-dominated scenario as for hadronic constraints. Under the constraint of $N_\gamma (>500\,\mathrm{GeV})\le 5000$, a large magnetic energy factor has to be invoked to make the SSC emission low, avoiding the violation of LHAASO observations. The leptonic contribution to the LHAASO detection number is normalized to $N_\gamma (>500\,\mathrm{GeV}) = 5000$ in Fig.~\ref{fig4} and consequently, the lower limits of $f_B$ are obtained and summarized in Table~\ref{tab:table3}. In addition, we also present the corresponding spectra as Fig.~\ref{fig7} for leptonic constraints in the time interval 200--300\,s with the same parameters as in Fig.~\ref{fig4}.

For a large bulk Lorentz factor $\Gamma$ (or a large dissipation radius), the internal $\gamma\gamma$ absorption inside the GRB jet becomes weak and the cutoff energy of the intrinsic SSC-cascade spectrum tends to be large, even extending to the LHAASO energy band. As a result, a large $f_B$ has to be involved in reducing the LHAASO detection number of VHE photons, e.g., $f_B \gtrsim 50$ for $\Gamma=1000$ and $f_B \gtrsim 150$ for $\Gamma=1500$, implying that a highly magnetized jet is required if a large bulk Lorentz factor is adopted. 


\begin{table}
\caption{\label{tab:table3}Constraints on the leptonic component under the assumption that the detection number by LHAASO is $\le 5000$ above 500\,GeV for the total contribution of two prompt time intervals.
}
\doublerulesep 0.1pt \tabcolsep 0.1pt
\begin{ruledtabular}
\begin{tabular}{ccc}
	\textrm{Descriptions}&\textrm{Symbols}&\textrm{Values}\\ 
        \colrule
			Bulk Lorentz factor 	&$\Gamma$  	& [500, 1000, 1500]  \\	
			Magnetic energy factor 	&$f_B$  & $\ge$[0.8, 50, 150] \\
            VHE photon number ($>10\,\rm TeV$) & $N_\gamma$  & $\le$[0, 0, $7\times 10^{-3}$]
\end{tabular}
\end{ruledtabular}
\end{table}

The expected detection number of $\gtrsim 10\,\rm TeV$ VHE photon is also listed in Table~\ref{tab:table3}. As can be seen in Fig.~\ref{fig4}, the detection number will be terminated at some critical energy, e.g., $\sim 3\,\rm TeV$ for $\Gamma=500$ and $\sim 4\,\rm TeV$ for $\Gamma=1000$ in the time interval $200-300\,\rm s$, and these sharp cutoffs are determined by the sharp cutoffs of electron distributions at the maximum electron energies. The maximum scattered photon energy of the IC process can not be larger than the initial electron energy in the Klein-Nishina regime given by Equation 2.50 in~\cite{RevModPhys.42.237}. Since the Fermi-LAT observation shows as an extra spectral component above 100\,MeV, the maximum synchrotron radiation energy in our numerical calculations is fixed to be a constant, i.e., $100\,\rm MeV$, and then one has $\Gamma \gamma_e^2 B \propto \mathrm{const.}$. Eventually, one has the maximum electron energy in the observed frame ${E_{e,\max }} = \Gamma {\gamma _e}{m_e}{c^2}/(1+z) \propto {\Gamma ^2}f_B^{ - 1/4} L_\gamma^{ - 1/4}$ with $B \propto {\Gamma ^{-3}}f_B^{ 1/2} L_\gamma^{ 1/2}$, which is almost the same with the maximum scattered photon energy. Therefore, in some cases, the detection number of $\gtrsim 10\,\rm TeV$ VHE photon could be zero if the maximum scattered photon energy is smaller than $10\,\rm TeV$. For two time intervals with the same bulk Lorentz factor $\Gamma$ and magnetic energy factor $f_B$, the critical cutoff energy difference is $10^{-1/4} \simeq 0.56$ since $L_\gamma=2\times 10^{53}\,\rm erg/s$ for 200--300\,s and $L_\gamma=2\times 10^{52}\,\rm erg/s$ for 300--400\,s are involved (see the difference between blue solid and blue dashed lines or red solid and red dashed lines in Fig.~\ref{fig4}). For the same time interval with the same $L_\gamma$, the critical cutoff energy difference is proportional to ${\Gamma ^2}f_B^{ - 1/4}$. These detection numbers for $> 10\,\rm TeV$ VHE photons listed in Table~\ref{tab:table3} could be larger if a larger maximum synchrotron radiation energy is taken. However, a too large maximum synchrotron radiation energy may violate the observations of Fermi-LAT above $100\,\rm MeV$. As a result, the $\gtrsim 10\,\rm TeV$ VHE photon may not originate from the leptonic scenario.

Note that we used the same spectral shape of keV/MeV radiations for two time intervals for both hadronic and leptonic constraints. The different photon spectral shapes of keV/MeV for the same luminosity would impact the results. For instance, for a softer low-energy photon index, i.e., smaller $\alpha_\gamma$, more photons will concentrate at lower energies, which will enhance the number density of the low-energy photon field and the  subsequent efficiencies of SSC scatterings and photo-hadronic interactions. As a result, the higher SSC-initiated and hadron-initiated cascade emissions can be expected for a smaller $\alpha_\gamma$, and then both leptonic and hadronic constraints will be more stringent  (i.e., larger $f_B$ and smaller $f_p$) when considering the LHAASO detection number $N_\gamma (>500\,\mathrm{GeV}) \le 5000$. While for a harder (larger) $\alpha_\gamma$, the constraints can be relaxed to some extent. In addition, the intrinsic spectral shape around TeV (spectral index $\alpha_{\rm TeV}$, without the EBL absorption but with the absorption inside the GRB jet) becomes softer for a softer $\alpha_\gamma$ and harder for a harder $\alpha_\gamma$ due to the internal $\gamma\gamma$ absorption. The overall cascade flux will become higher for a softer $\alpha_\gamma$ and lower for a harder $\alpha_\gamma$ since the effect of radiation efficiency is much more dominant. However, the detection number of $>10\,\rm TeV$ photons will be higher for a harder $\alpha_{\rm TeV}$ (or $\alpha_\gamma$) and lower for a softer $\alpha_{\rm TeV}$ (or $\alpha_\gamma$) when one normalizes the LHAASO detection number to $N_\gamma (>500\,\mathrm{GeV}) = 5000$. Besides, for the leptonic constraints, the cascade emission initiated by the SSC photons is dominated by the unabsorbed SSC photons that keep a similar spectral shape to the synchrotron radiation, and thus the spectral shape of GeV-TeV photons would change with the spectral shape of keV/MeV radiations. However, this effect can be neglected for the hadronic constraints since the hadron-initiated EM cascade can be fully developed and show a universal spectral shape.

\section{Discussions and Conclusions}\label{sec4}
 GRB 221009A is the most luminous GRB detected ever. The abundant observations of GRB 221009A, including keV/MeV, GeV/TeV EM radiations, and the neutrino upper limit, provide us with a unique opportunity to explore the origin of VHE gamma-rays in the prompt emission phase. In this work, combining the multi-wavelength and multi-messenger observations, we have studied the origins of VHE gamma-rays in the prompt emission of GRB 221009A, including the leptonic and hadronic origins, as well as the consequent constraints on them. We find the required baryonic loading factor is $f_p\lesssim 2$ for a large range of bulk Lorentz factor. The VHE and $>10\,\rm TeV$ photons can originate from the hadronic processes with a detection number of $\gtrsim10 \,\rm TeV$ photon around unity in the GRB prompt emission phase. In addition, the magnetic energy factor should be large to match the LHAASO observations, especially for a large bulk Lorentz factor, implying a highly magnetized jet and supporting the Blandford $\&$ Znajek (BZ) mechanism as the possible central engine model \citep{1977MNRAS.179..433B,2017ApJ...849...47L}. The highly magnetized jet may induce strong magnetic dissipation undergoing an efficient magnetic-to-kinetic energy conversion and the released energy can be distributed to electrons and protons through the magnetic reconnection acceleration and the possible accompanying turbulence acceleration~\citep{1994MNRAS.270..480T,2009MNRAS.394.1182K,2011ApJ...726...90Z}, although the internal shock scenario can still operate but may be in an inefficient acceleration situation~\citep{2011ApJ...726...75S,2011ApJ...726...90Z}. Moreover, our results suggest that the SSC process can contribute to sub-TeV photons but may not produce enough number of $\gtrsim10 \,\rm TeV$ photons in the prompt emission phase. 

 The constraints are obtained based on the detection number of VHE photons by LHAASO $N_\gamma(> 500\,\rm \mathrm{GeV})\le 5000$, the detection number of high-energy neutrinos $N_\nu\le 3$, and the gamma-ray emission at Fermi-LAT energy band (100\,MeV--300\,GeV) less than the Fermi-LAT observations. Our constraints on the microscopic physical parameters are conservative considering the possible presence of radiations from external shock, external IC due to the possible external photon field, and synchrotron of intermediated particles such as charged pions and muons from the photomeson production process. We implemented separately hadronic constraints and leptonic constraints, each of both should be satisfied with the observational limitations. Therefore, our results are conservative considering the possible contribution of another component. During the leptonic constraints, the hadronic component can be easily neglected (by setting $f_p=0$), and during the hadronic constraints, the SSC component is set as zero by hand. Although during the leptonic constraints, a high $f_B$ has been derived, a typical $f_B=1$ is adopted  for the hadronic constraints since the impact of $f_B$ on the hadron-initiated EM cascade is weak so that we can implement relatively independent constraints on $f_p$ during the hadronic constraints (see Section~\ref{sec:hadconstraints} for details).

LHAASO measurement makes the GRB 221009A the first GRB with the detection of photons above 10 TeV. The expected VHE photon number is relevant to the adopted EBL model. We also tried different EBL models and found the effects of different EBL models on the detection of $\sim 500\,\rm GeV$ photons are quite weak since the optical depths of different EBL models at $500\,\rm GeV$ are almost same. For the 500 GeV photons, the threshold energy for the pair production is $\simeq 0.5\,\rm eV$ (corresponding to the EBL wavelength of $\simeq 2.5\,\rm \mu m$), where the constraint on the EBL model is tight and the difference of different EBL models is small (see, e.g., \cite{2021MNRAS.507.5144S}). The detection number at $\sim 500\,\rm GeV$ determines the total detection number by LHAASO $N_\gamma(\gtrsim 500\,\rm \mathrm{GeV})$. As a result, different EBL models will not affect our constraints significantly. However, for photons with energies above $10\,\rm TeV$, if a weaker EBL model is involved, more $>10\,\rm TeV$ photons will be expected, and vice versa. In this work, a recent EBL model given by \cite{2021MNRAS.507.5144S} is adopted for numerical calculations, which is a relatively strong EBL model (see, e.g., \cite{2022arXiv221007172B}). Besides, we tried the relatively weak EBL model described by \cite{2010ApJ...712..238F}, the detection number of $>10\,\rm TeV$ photons will increase by a factor of $\sim 2$. Considering the uncertainties of the EBL given by \cite{2021MNRAS.507.5144S}, which almost covers the uncertain region of most of EBL models, the change of the detection number of $>10\,\rm TeV$ photons ranges with a factor of $\sim 0.2-4.5$.

In the future, once detailed information on LHAASO observations can be available, e.g., the early detection of VHE photons during the prompt emission phase and the coincidence of temporal variability between the VHE photons and the keV/MeV radiations (the behavior as in \cite{2017ApJ...844...56T} for high-energy gamma-rays and keV/MeV radiations), the VHE photons (at least partial VHE photons) will tend to support the internal origin. The precise VHE photon number originating from the prompt emission phase will provide more stringent constraints for our model. In addition, the detailed spectral shape of LHAASO observations, combined with the Fermi-LAT observations and the observed keV/MeV emissions, can be used to be implemented the detailed multi-wavelength spectral fitting. Therefore, the precise contribution of each component (leptonic or hadronic) can be studied and then more stringent constraints on the parameters can be expected.

The absorbed VHE photons by EBL can initiate the intergalactic EM cascade, generating the angle-extended and time-delayed GeV emission due to the deflections of electron pairs in the intergalactic magnetic field (IGMF) \citep{2021Univ....7..223A}. The detection of such a time-delayed GeV emission usually needs a weak IGMF to generate an enough high GeV flux \citep{2009PhRvD..80l3012N}. Non-detection of the delayed GeV emission will exclude the possibility of extremely weak magnetic fields. Such an approach has been used to provide the lower bound of the IGMF based on the blazar observations~\citep{2002ApJ...580L...7D,2015ApJ...814...20F,2015RAA....15.2173Y} and the GRB observations~\citep{PhysRevD.101.083004}. GRB 221009A can be a unique source to study the intergalactic gamma-ray propagation and constrain the intergalactic environment.

Although the external origin in the afterglow phase, the possible ALPs scenario, and the EM cascade in the extragalactic medium initiated by UHECRs may (partially) operate to be responsible for the VHE and $>10\,\rm TeV$ photons in GRB 221009A, we emphasize the detection of VHE and even $>10\,\rm TeV$ photons in the prompt emission phase is plausible without involving exotic physics and can provide strong constraints on the GRB properties. In the future, more GRBs detected at the VHE energy band by LHAASO and the Cherenkov Telescope Array \citep{2013APh....43..252I} can help us understand particle accelerations, the jet composition, and radiation mechanisms in the prompt emission phase.

\acknowledgments

We thank Prof. Wei-Hua Lei for the helpful discussion. This work is supported by the National Natural Science Foundation of China under grants No.12003007, U2031105, U1931201, and U1931203, the Fundamental Research Funds for the Central Universities (No. 2020kfyXJJS039) and the China Manned Space Project (CMS-CSST-2021-B11).

\bibliography{reference}{}
\bibliographystyle{aasjournal}

\end{document}